\documentclass[prl,rd,showpacs,twocolumn,superscriptaddress]{revtex4-1}
\usepackage{amsfonts,amsmath,amssymb,amsthm}
\usepackage[bookmarks, bookmarksopen, bookmarksnumbered]{hyperref}

\usepackage{graphicx}
\usepackage{color}
\usepackage{siunitx}
\usepackage[utf8]{inputenc}
\usepackage[T1]{fontenc}
\usepackage{lmodern}
\usepackage[right]{rotating}
\usepackage{longtable}
\usepackage{array}
\usepackage{multirow}
\usepackage{paralist}
\usepackage{colortbl}

\usepackage{aas_macros}

\renewcommand{\i}{\mathrm{i}}

\usepackage{color, colortbl}
\definecolor{Gray}{gray}{0.9}

\graphicspath{{figs/}}

\usepackage[normalem]{ulem}

\begin{document}

\title{Universal relations for gravitational-wave asteroseismology of proto-neutron stars}
\date{\today}

\author{Alejandro \surname{Torres-Forn\'e}}
\affiliation{Max Planck Institute for Gravitational Physics 
	(Albert Einstein Institute), D-14476 Potsdam-Golm, Germany}

\author{Pablo \surname{Cerd\'a-Dur\'an}}
\affiliation{Departamento de
	Astronom\'{\i}a y Astrof\'{\i}sica, Universitat de Val\`encia,
	Dr. Moliner 50, 46100, Burjassot, Spain}

\author{Martin \surname{Obergaulinger}}
\affiliation{Departamento de
	Astronom\'{\i}a y Astrof\'{\i}sica, Universitat de Val\`encia,
	Dr. Moliner 50, 46100, Burjassot, Spain}
\affiliation{Institut fur Kernphysik, Theoriezentrum, Schlossgartenstr. 2, D-64289 Darmstadt, Germany}

\author{Bernhard \surname{M{\"u}ller}}
\affiliation{Monash Centre for Astrophysics, School of Physics and Astronomy, Monash University, VIC 3800, Australia}

\author{Jos\'e A.~\surname{Font}}
\affiliation{Departamento de
 Astronom\'{\i}a y Astrof\'{\i}sica, Universitat de Val\`encia,
 Dr. Moliner 50, 46100, Burjassot, Spain}
\affiliation{Observatori Astron\`omic, Universitat de Val\`encia, 
 Jos\'e Beltr\'an 2, 46980, Paterna, Spain}

\begin{abstract}
State-of-the-art numerical simulations of core-collapse supernovae reveal that the main source of gravitational waves is the excitation
of proto-neutron star modes during post-bounce evolution. In this work we derive universal relations that relate the frequencies of the most common oscillation modes observed, i.e.~g-modes, p-modes and the f-mode, with fundamental properties of the system, such as the surface gravity of the proto-neutron star or the mean density in the region enclosed by the shock. These relations are independent of the equation of state, the neutrino treatment, and the progenitor mass and hence can be used to build methods to infer proto-neutron star properties from gravitational-wave observations alone. We outline how these measurements could be done and the constraints that could be placed on the proto-neutron star properties. 
\end{abstract}

\LTcapwidth=\columnwidth


\maketitle


Core-collapse supernova (CCSN) explosions are a promising source of gravitational waves (GW) and might be one of the next
discoveries of current or future ground-based GW observatories. The most common CCSN type, a neutrino-driven explosion, is expected to be observable with Advance LIGO and Virgo within our galaxy \cite{Gossan:2016} at a rate of about three per century \cite{Adams:2013}. CCSN events mark the end of the life of massive stars ($8$-$100$~$M_\odot$) following the formation of a heavy iron core that collapses under its own gravity. The end-result is the formation of a proto-neutron star (PNS) at densities above nuclear-matter density with a radius of $\sim30$~km and a stalled accretion shock at $\sim100$~km. This situation is maintained for $0.1$-$2$~s, while accretion proceeds onto the PNS as it cools down through neutrino emission. The shock-PNS system may suffer instabilities, in particular convection and the standing accretion shock instability (SASI). Those instabilities are crucial for shock revival, as they allow for an enhanced energy deposition in the post-shock region by the neutrinos streaming out of the PNS. At the same time these instabilities break spherical symmetry and induce perturbations of the PNS that produce a rich, stochastic GW signal that lasts until the onset of explosion, if successful, or up to the formation of a black hole. Detailed descriptions of the CCSN explosion mechanism and GW emission can be found, e.g., in \cite{Janka:2017} and \cite{Kotake:2017}, respectively. 

\begin{figure*}[t!]
	\includegraphics[width=0.49\textwidth]{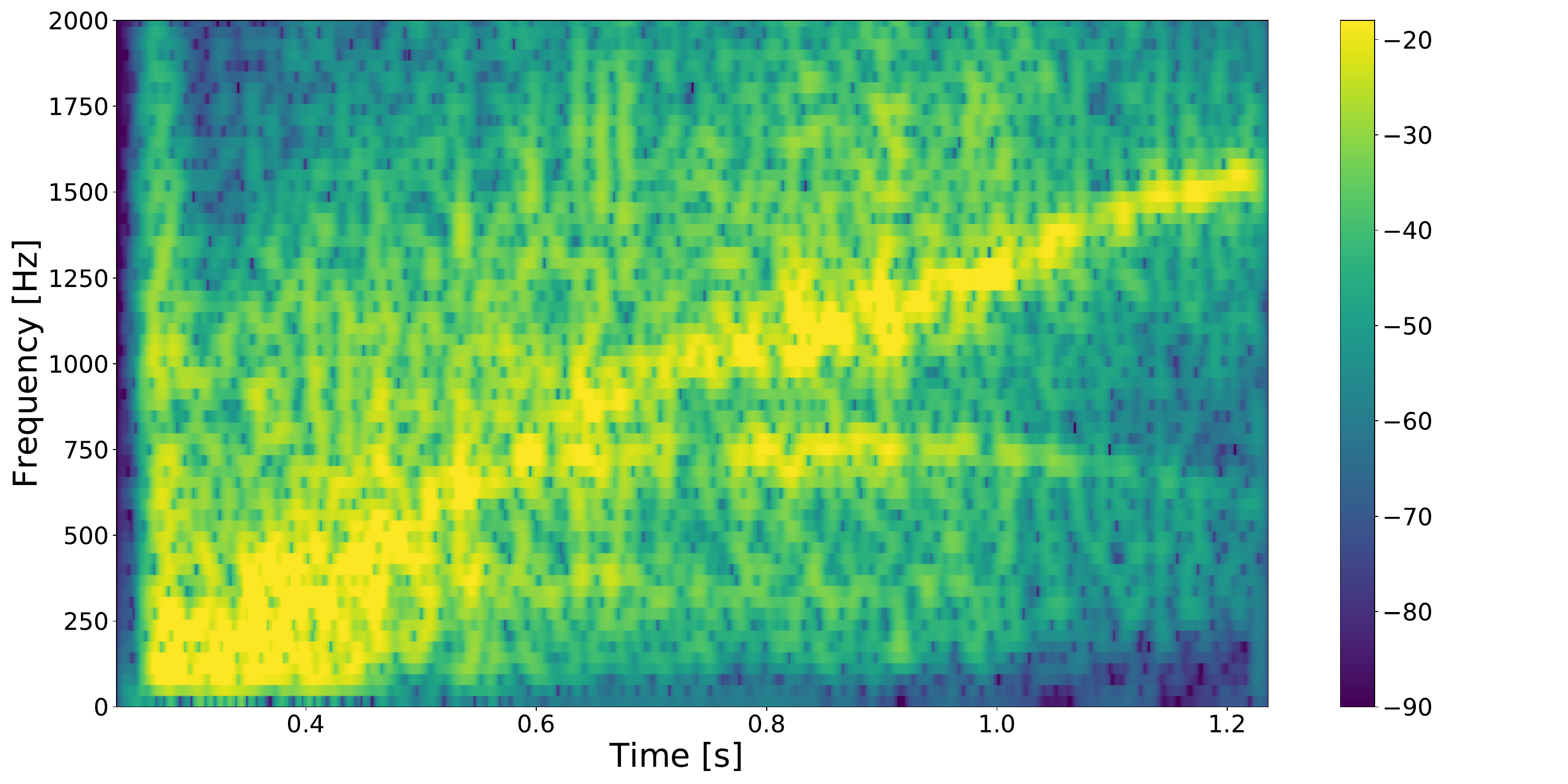}
	\includegraphics[width=0.49\textwidth]{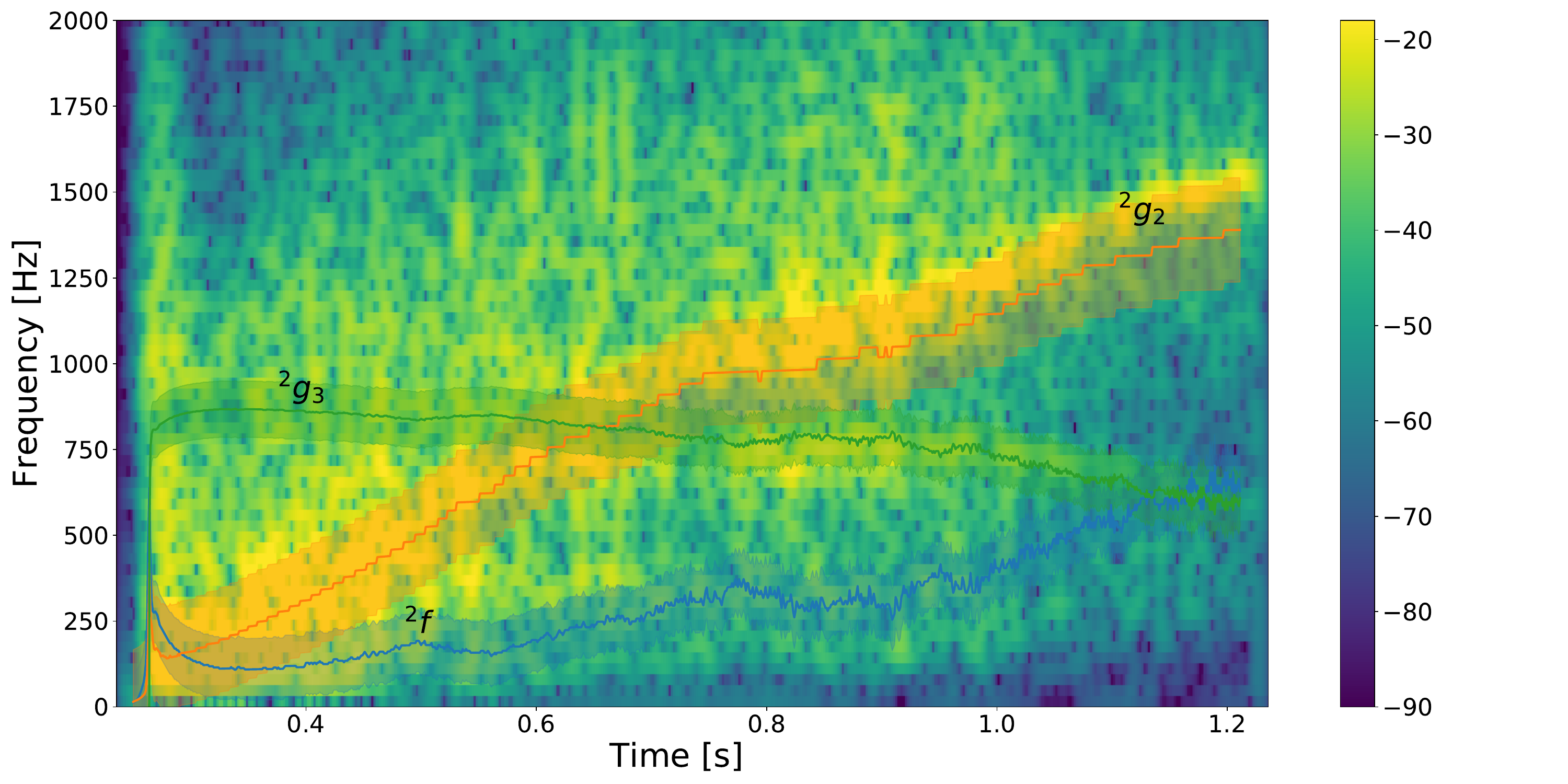}
	\vspace{-0.45cm}
	\caption{Left: Time-frequency representation (spectrogram) of a GW from the collapse of a $20$~$M_\odot$
	progenitor (model s20 in \cite{TF:2019}). The two main features (arches with a power excess) were identified as g-modes. Right:   predicted frequencies for several modes with 2$\sigma$ errors (shaded regions) using the universal relations.}
	\label{fig:spect}
\end{figure*}

Numerical simulations have shown that the GW signal, albeit highly stochastic, displays some clear trends in the time-frequency plane (spectrograms) as in the example shown in the left panel of Fig~\ref{fig:spect}. This is the result of the excitation of PNS g-modes and, in some cases, of SASI modes \cite{Murphy:2009,Cerda-Duran:2013,Mueller:2013,Yakunin:2015,Kuroda:2016,Andresen:2017,Powell:2018,Radice:2018}. Recent work \cite{TF:2018,Morozova:2018,TF:2019} has established that the features observed in the GW spectrograms can be very accurately matched to the $l=2$ PNS eigenmodes, $l$ being the order of the spherical-harmonic decomposition. Those are obtained from the solution of an eigenvalue problem at each time-step of the simulation, using angular-averaged profiles of the simulation data as a background. Particularly important for an accurate matching has been the inclusion of general relativity and space-time perturbations \cite{TF:2019}. This analysis enables a detailed study of the behaviour of CCSN GW signals in spectrograms without the need of performing computationally challenging multidimensional simulations. Instead, affordable spherically-symmetric simulations serve the purpose, supplemented with the computation of the eigenfrequencies for $l=2$ perturbations, responsible for the GW emission.
Those simulations, however, do not allow to compute the amplitude of the waveform but only its frequency evolution. In this {\it Letter} we show that at each instant during the PNS evolution, the characteristic frequency for the different modes does not depend on 
the exact structure of the PNS but can be estimated from the general properties of the remnant. Furthermore, the relations we derive are {\it universal}, as they do not depend on the equation of state (EoS), progenitor star, or neutrino treatment. Therefore, they provide the potential to be used as a basis for parameter inference algorithms once GW observations from CCSN become available. 

\begin{table*}[t!]
	\centering
	\caption{Parameters of the universal relations found for each of the modes considered.
		Fits are of the form $f= a+b x+cx^2+dx^3$, with $f$ expressed in Hz, and using $M_\odot$ and km
		as units for mass and length in the quantities expressed in the variable $x$, except for $\rho_{C}$ and
		$p_{\rm C}$, expressed in cm ($G=c=1$). 
		Note that some coefficients are not used in some fits.
		$R^2$ and $\sigma$ are the 
		correlation coefficient and the standard deviation of the data (in Hz) with respect to the fit, respectively.
		}
		\begin{tabular}{lccccccc}
		\hline\hline
		mode& $x$ & a & $b /10^5$ & $c / 10^6$ & $d / 10^{9}$ &$R^2$ & $\sigma$ \\ 
		\hline 
		$^2f$	& $\sqrt{M_{\rm{shock}}/R_{\rm{shock}}^3}$& - &$ 1.410 \pm 0.004 $ & $-4.23 \pm 0.06$ & -
		& 
		0.966 & 45  \\ 

		$^2p_1$	& $\sqrt{M_{\rm{shock}}/R_{\rm{shock}}^3}$& -&$2.205\pm 0.007$ &$4.63 \pm 0.09$&-
		 & 
		 0.991 & 61 \\

		$^2p_2$	& $\sqrt{M_{\rm{shock}}/R_{\rm{shock}}^3}$& -&$ 4.02 \pm 0.02$ &$7.4 \pm 0.3$  &-
		& 
		 0.983 & 123 \\

		$^2p_3$	& $\sqrt{M_{\rm{shock}}/R_{\rm{shock}}^3}$& -& $6.21 \pm 0.03$&$ -1.9 \pm 0.6$ &-
		& 
		0.979 & 142 \\ 
		
		$^2g_1$	& ${M_{\rm{pns}}/R_{\rm{pns}}^2}$&- & $8.67 \pm 0.03$ & $ -51.9 \pm 0.5 $ & -
		& 
		0.958 & 205 \\ 

		$^2g_2$	& ${M_{\rm{pns}}/R_{\rm{pns}}^2}$&- & $5.88 \pm 0.03$& $-86.2 \pm 1.0$ 
		& $4.67 \pm 0.08$ &
		 0.956 & 85 \\ 

		$^2g_3$	& $\sqrt{M_{\rm{shock}}/R_{\rm{shock}}^3}~p_C/\rho_C^{2.5}$ & $905 \pm 3$ & $-79.9 \pm 1.7$&$-11000 \pm 2000$ & -
		&
		0.925 & 41 \\ 
		\hline \hline
	\end{tabular} 
	\label{tab:fits}
\end{table*}


Our analysis is based on $25$ 1D simulations with different combinations of numerical codes, gravity approximations, EoS, and progenitor stars. The simulations are performed with the \texttt{AENUS-ALCAR} code \cite{Obergaulinger:phd,Just:2015} and the \texttt{CoCoNuT} code \cite{Dimmelmeier:2005}. Both are multidimensional Godunov-based Eulerian hydrodynamics codes for spherical polar coordinates and include neutrino-transport.Both codes include multigroup treatment for three neutrino species (electron neutrinos and antineutrinos, and heavy flavor neutrinos). \texttt{AENUS-ALCAR} uses an algebraic Eddington-factor method with an $M_1$ closure \cite{Just:2015}, whereas \texttt{CoCoNuT} uses a less sophisticated approach \cite{Mueller:2015} based on a stationary transport solution with only a one-moment closure. In both cases neutrino interactions include charged-current reactions of electron neutrinos and antineutrinos with nucleons and nuclei, neutrino scattering off nucleons and nuclei, and neutrino production by nucleon bremsstrahlung. \texttt{AENUS-ALCAR} also includes electron-positron pair processes and inelastic scattering off electrons. Regarding the gravity treatment, \texttt{AENUS-ALCAR} can use either Newtonian gravity or a pseudo-relativistic potential (TOV-A in \cite{Marek:2006}) while \texttt{CoCoNuT} uses general relativity in the XCFC formulation \cite{CC:2009}. 
We use 6 different finite-temperature EoS in our simulations: LS220 \cite{Lattimer:1991}, GShen-NL3 \cite{GShen:2011}, HShen \cite{HShen:2011}, SFHo\cite{Steiner:2013}, BHB-$\Lambda$ \cite{Banik:2014} and HShen-$\Lambda$ \cite{HShen:2011}, the latter two including $\Lambda$ hyperons. However, hyperons are not included in the neutrino interaction rates. As initial data we use a selection of pre-supernova models from \cite{Woosley:2002} including solar-metallicity models between $11$ and $75$~$M_\odot$ and one $20$~$M_\odot$ model with $10^{-4}$ solar metallicity. The complete list of simulations can be found in the Supplementary Material.
For each simulation and at each time after bounce, the eigenmode frequencies of the region including the PNS and the shock are determined using the code \texttt{GREAT} \cite{TF:2019} which includes corrections for space-time perturbations of the lapse and conformal factor in general relativity. This eigenmode calculation does not provide a classification of the modes by itself. We use the 
procedure described in \cite{TF:2019} to classify modes in f-modes ($^lf$), p-modes ($^lp_n$) and g-modes ($^lg_n$), where 
$n$ indicates the number of radial nodes. We restrict ourselves to $l=2$ modes, which are the dominant ones for GW emission.
Compared to our mode classification presented in \cite{TF:2019}, we here re-label some previously misclassified modes: one of the p-modes is reclassified as $^2g_1$, for all other g-modes $n$ is increased by one ($^2g_n \to$ $^2g_{n+1}$), and for all p-modes above the reclassified p-mode $n$ is decreased by one ($^2p_{n+1} \to$ $^2p_{n}$). The new and wider set of simulations used in this work (\cite{TF:2019} only considered 2 simulations) shows that the lowest order g-mode had been misclassified as a p-mode. This is supported by three aspects of the simulations: the energy density distribution of the mode, much more concentrated in the PNS interior, the time evolution of the frequency, which often presents crossings with other p-modes, and the behaviour of the universal relations found in this work (see below).


\begin{figure*}
\includegraphics[width=0.49\textwidth]{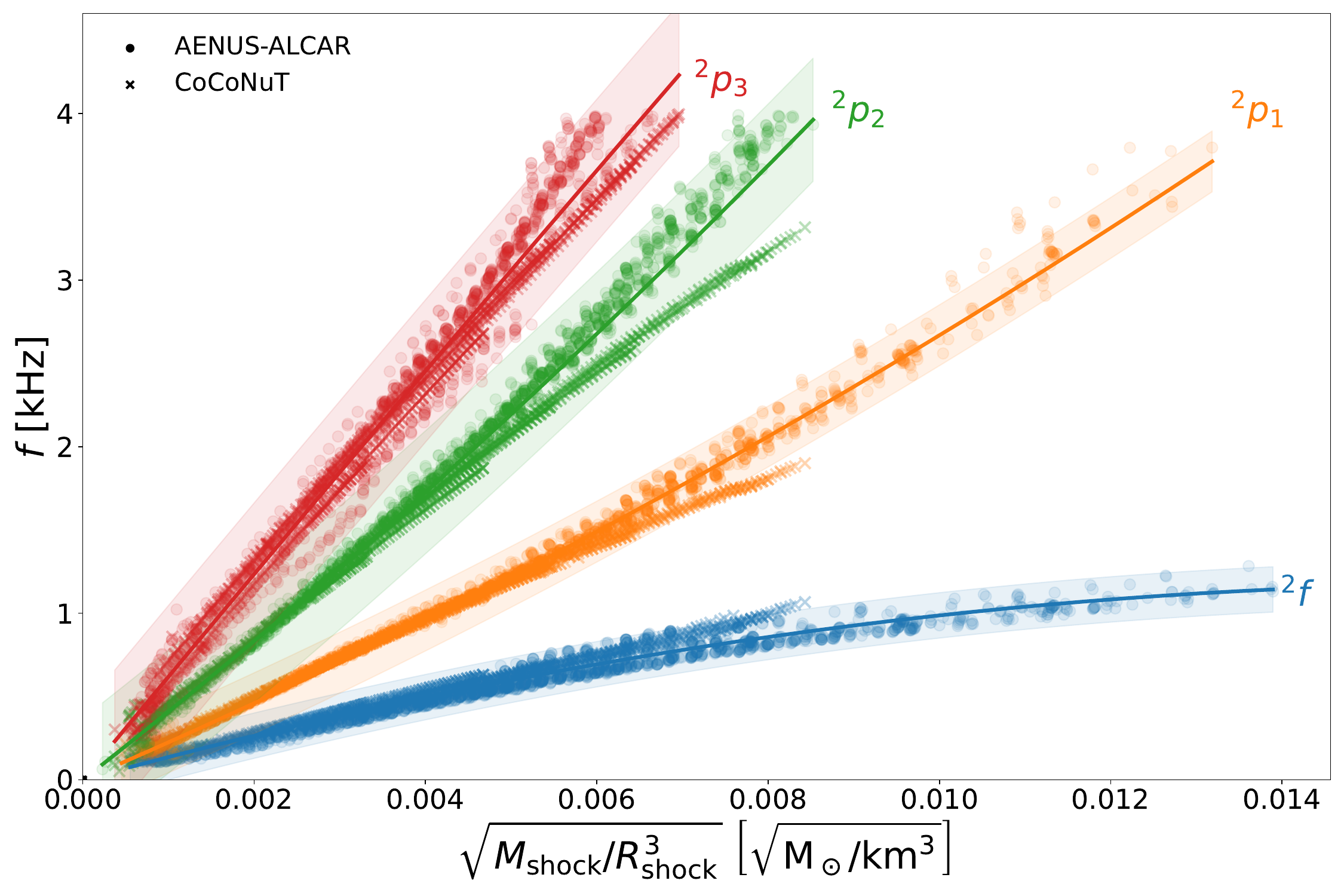}
\includegraphics[width=0.49\textwidth]{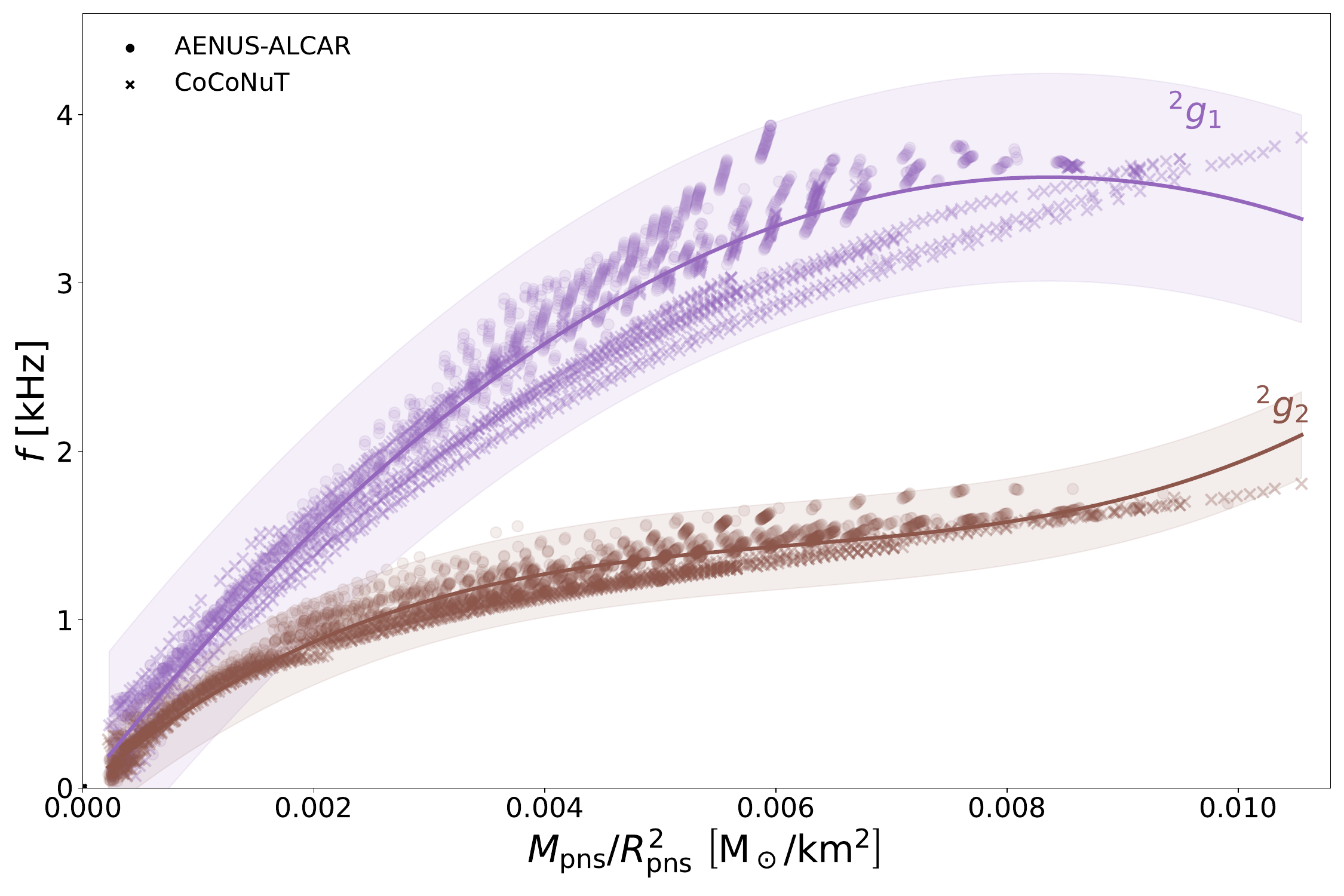}
\vspace{-0.4cm}
\caption{Fits of the different modes. The left panel shows the f-mode and the first three p-modes while the right panel shows the first two g-modes. The results from \texttt{AENUS-ALCAR} and \texttt{CoCoNuT} are represented with solid circles and crosses, respectively. Shaded areas indicate 2$\sigma$ error intervals. }
\label{fig:mode_fits}
\end{figure*}

We focus on the f-mode and on the lowest-order g- and p-modes, which have been shown to be the only ones observable in the GW spectrum \cite{TF:2018,TF:2019}. We obtain the eigenfrequency of each mode considering all evolution times and all simulations as a single dataset and try to find relations between this frequency and the properties of the system, namely the PNS mass and radius 
($M_{\rm PNS}$ and $R_{\rm PNS}$), the shock radius ($R_{\rm shock}$), the total mass inside the shock ($M_{\rm shock}$), the
central density and pressure ($\rho_{\rm c}$ and $P_{\rm c}$), as well as different thermodynamical quantities 
at different radii. To find the best possible relations we systematically perform fits of the eigenfrequencies with polynomials of the form 
$f=a + b x + c x^2 + d x^3$,
with $x=A^\alpha B^\beta C^\gamma D^\delta$ and where $A, \cdots, D$ are all different combinations of the quantities defining the system
and $\alpha, \cdots,\delta$ are exponents ranging in the interval [-3,3] in steps of $0.5$. Universal relations for each of the modes are built
using the best-fitting combinations together with our intuition on the physical processes that should determine the frequencies. The results for the fits and the combination used for $x$ are presented in Table \ref{tab:fits}. 
For most of the fits we have considered $a=0$, i.e. imposed that the frequency goes to zero as $x$ vanishes. $d\ne 0$ was considered only in one case since in all others this coefficient did not improve the fit significantly.

Fig.~\ref{fig:mode_fits} shows the fits for the universal relations. In each case we observe a scatter of the data around the best fit, that can be characterised by a standard deviation $\sigma$ (given in Table~\ref{tab:fits}). This scatter does not depend systematically on the EoS or on the progenitor used for the simulation. The only systematic behaviour found is caused by employing two different codes (indicated with different symbols in the plots) which leads to slightly different relations at the upper frequency part of the fits. This systematic difference may be due to the different gravity treatment in either code (better in \texttt{CoCoNuT}) or to the different neutrino treatment (better in \texttt{AENUS-ALCAR}). Although the errors in the fits are fairly small, the observed systematics indicates that they could be decreased even further by performing more complete numerical simulations, which might be necessary for the purpose of inference (as described below). 
 
The lowest order g-modes, $^2g_1$ and $^2g_2$, depend primarily on the surface gravity of the PNS, i.e.~$M_{\rm PNS}/R^2_{\rm PNS}$. The reason is because buoyancy inside the PNS is responsible for the excitation of these modes and the phase velocity of the associated (gravity) waves depends directly on the surface gravity. Measuring the frequency of these modes gives an idea of the compactness of the PNS. 
 
The f-mode and p-modes depend on the square root of the mean density inside the shock, i.e.~$\sqrt{M_{\rm shock}/R_{\rm shock}^3}$, as their frequency is primarily determined by the local sound speed and the size of the region containing the modes. The frequency of these modes track primarily the location of the shock. The general behaviour matches the dependence seen in previous work \cite{Mueller:2013} and is expected according to general theory of modes in stars \cite{Andersson:1998}. We note that the $^2g_1$ mode, improperly labeled as a p-mode in \cite{TF:2019}, behaves as a g-mode, since its frequency depends on the PNS surface gravity. Attempts to fit it with a p-mode dependence result in very poor fits. This is further evidence that, despite of its high frequency, this mode is indeed a g-mode.

The $^2g_3$ mode presents a significantly different behaviour to other g-modes. Our analysis shows that, despite of being classifies as a g-mode, it cannot 
be fitted using the PNS surface gravity ($R^2=0.76$). Instead, in the best fit (see table~\ref{tab:fits}), the dominant behaviour is related to the
mean density inside the shock (similarly to p-modes) but additional corrections including central density and pressure are necessary. Without this additional correction the quality of the 
fit degradates significantly ($R=0.85$).
The term $p_c/\rho_c^{2.5}$ is very close to $p_c/\rho_c^{\Gamma_1}$, which is related to the entropy (here $\Gamma_1$ is the adiabatic index). However, neither using the entropy itself nor $\Gamma_1$ instead of $2.5$,
improves the fit. Note that the frequency of this mode decreases with increasing $x$, since $b<0$. In fact the time evolution of the frequency is monotonically decreasing in all our simulations. The puzzling behaviour of this mode may require special attention, mainly because it has been observed in the GW spectrograms of several numerical simulations \cite{Kawahara:2018,TF:2019}.

From these results we conclude that it is possible to derive relations that connect the frequencies of the PNS modes, observable in the GW signal, with intrinsic properties of the PNS. These relations are universal, within some uncertainties, since they do not depend strongly on the EoS, neutrino treatment, or progenitor star. Therefore, they could be used for parameter estimation when GW observations
 of CCSN are accomplished. This idea was first proposed by \cite{Andersson:1998} for cold neutron stars but with the newly proposed universal relations we show its feasibility in the CCSN context.
 The procedure that could be used for inference is outline next. First, the traces of the different modes have to be extracted from the spectrograms. The dominant emission mode, as observed in all previous work \cite{TF:2018,Morozova:2018,TF:2019}, would correspond to the $^2g_2$ mode. If present, the pattern of the $^2g_3$ mode is also easy to recognise, as it monotonically decreases with time. Finally, in the presence of the SASI, the f-mode can be easily identified due to its low frequency. Other modes that may appear will likely have lower amplitudes and could be classified as higher or lower overtones of those already identified. Each identified mode follows a track in the time-frequency plane, $f(t)$, with an error associated with the measurement process of a GW signal buried in detector noise. Each of those functions $f(t)$ can be transformed into a $x(t)$ using the adequate universal relations $f(x)$. The error for $x(t)$ will depend on both the error for $f(t)$ and the error associated with the universal relation $f(x)$. As a result, this method will allow to measure 
 $M_{\rm PNS}/R_{\rm PNS}^2$ from the dominant g-modes and $\sqrt{M_{\rm shock}/R_{\rm shock}^3}$ from the f-mode, with their uncertainties. The development of an inference pipeline based on this proposal is one of our main immediate goals and will be discussed elsewhere.


Instead of providing an example of inference, i.e.~the solution of the inverse problem, out of the scope of this work, we present here an example of the direct problem. We use the 2D CCSN simulation s20 of \cite{TF:2019} and using only the values of the $M_{\rm PNS}/R_{\rm PNS}^2$ and $\sqrt{M_{\rm shock}/R_{\rm shock}^3}$ extracted from the simulation, we compute the predicted frequencies
 for the dominant modes. The right panel of Fig.~\ref{fig:spect} compares those frequencies with the GW spectrogram of the simulation and shows that both agree within the errors of the universal relations. Note that data from the s20 simulation was not included in the models used to build the universal relations, which are all 1D. This is an indication that our relations are valid even when applied to multidimensional models and therefore are truly universal.

\begin{figure}
	\includegraphics[width=0.49\textwidth]{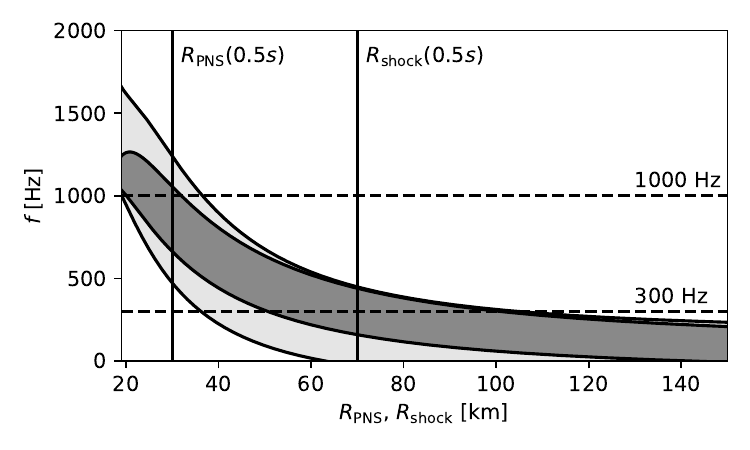}
	\caption{Range of frequencies for the $^2g_2$ (light shade) and the $^2f$ mode (dark shade) as a function of the PNS and shock radius, respectively, 
	for masses in the interval $1.2-2.5 ~M_{\odot}$ considering $2\sigma$ errors from the universal relations. Vertical lines show the PNS and shock radius $0.5$~s
	post-bounce and horizontal dashed lines the frequency of the $^2g_2$ mode ($\sim 1000$~Hz) and the $^2f$ mode ($\sim 300$~Hz) at the same time.}
	\label{fig:radius}
\end{figure}

The relations presented in this work may eventually allow to measure physical properties of a PNS during its first second of life, out of purely GW information. The measurement of the PNS surface gravity gives information about the EoS of nuclear matter at finite temperature, which has not been probed yet in other astrophysical scenarios (let alone at the laboratory). The only similar scenario is the post-merger evolution of binary neutron stars but the typical frequencies are somewhat higher and thus less accessible with current GW detectors~\cite{DePietri:2018}. Measuring $M_{\rm PNS}/R_{\rm PNS}^2$ also allows to constraint the PNS radius. The minimum mass of the iron core at the onset of collapse is set by the Chandrasekhar mass \cite{Chandrasekhar:1938} for a gas of degenerate electrons, $M_{\rm Ch0}=5.83 Y_e^2$, where $Y_e \simeq 0.46$ is the electron fraction for iron, which results in a minimum mass of $1.2 M_\odot$. The maximum mass depends on finite temperature effects \cite{Woosley:2002}. Stellar evolution models show that it can be as high as $\sim 2.5 M_\odot$, although for solar-metallicity models it does not exceed $2 M_\odot$ \cite{Woosley:2002}. The iron core collapses into the PNS in a typical timescale of a few $100$~ms. After this the mass accretion rate drops significantly due to the lower density of the outer layers. Therefore, at about $0.5$~s after bounce the mass of the PNS is very close to the mass of the iron core and is in the range $1.2-2.5 M_\odot$. Similar arguments hold for $M_{\rm shock}$, which is approximately equal to $M_{\rm PNS}$, because of the small amount of mass between the PNS and the shock. This allows to transform the universal relations into constraints for the shock and PNS radii. Fig.~\ref{fig:radius} shows an example of how to constraint the PNS radius by measuring the frequency of the dominant $^2g_2$ mode. For the s20 model the frequency at $0.5$~s post-bounce is about $1$~kHz, which would correspond to a measurement of the PNS radius in the $2\sigma$ confidence interval $19 - 37$~km. The PNS radius computed from the actual simulation is $30$~km, which falls within this interval. Similarly, a measurement of the f-mode frequency of about $300$~Hz places a constraint on the shock radius in the $2\sigma$ interval $50-100$~km, at $0.5$~s post-bounce to be compared with the value of $70$~km obtained from the simulation. In an actual CCSN event there could be additional constraints for the mass of the iron core, which would reduce further the errors in the estimation, e.g.~constraints from the neutrino luminosity and spectrum, from the observation of the progenitor star and from the direct observation of the compact remanent, possibly as a pulsar, decades after the explosion.

\acknowledgments

\bigskip

We thank Marie Anne Bizouard and Sanjay Reddy for useful discussions. Work supported by the Spanish MINECO (grant AYA2015-66899-C2-1-P), by the Generalitat Valenciana (PROMETEOII-2014-069) and by the European Gravitational Observatory (EGO-DIR-51-2017). PCD acknowledges the support from the Ramon y Cajal program of the Spanish MINECO (RYC-2015-19074). JAF acknowledges support from the European Union's Horizon 2020 RISE programme H2020-MSCA-RISE-2017 Grant No.~FunFiCO-777740. 
MO acknowledges support from the European Research Council under grant EUROPIUM-667912, and from the Deutsche Forschungsgemeinschaft through Sonderforschungsbereich SFB 1245 "Nuclei: From fundamental interactions to structure and stars". BM acknowledges support by ARC grant FT160100035.

\bibliographystyle{apsrev4-1}
\bibliography{references}

\end{document}